\documentstyle [12pt,twoside, epsf]{article}

\def\picill#1by#2(#3)
{\vbox to #2
{\hrule width #1 height 0pt depth 0pt
\vfill\epsffile{#3}}}

\begin{document}

\date{}

\title{\bf Teleportation Topology}

\author{Louis
H. Kauffman\\ Department of Mathematics, Statistics \\ and Computer Science (m/c
249)    \\ 851 South Morgan Street   \\ University of Illinois at Chicago\\
Chicago, Illinois 60607-7045\\ $<$kauffman@uic.edu$>$}

 \maketitle
  
 \thispagestyle{empty}
 
 \subsection*{\centering Abstract}

{\em
The paper discusses teleportation in the context of comparing quantum and topological points of view. }
 
\section{Introduction}

We discuss the structure of teleportation. By associating matrices to the preparation and measurement states, we show that for unitary
$M$ there is a teleportation procedure for obtaining $M|\psi \rangle$ from a given state $|\psi \rangle.$ 
The key to this construction is a diagrammatic intepretation of matrix multiplication. This interpretation applies equally well to a topological
composition of a maximum and a minimum that underlies the structure of the teleportation. Thus we obtain a novel view of the structure of 
teleportation. 
\bigbreak

This paper is a preliminary report
on joint work with H. Carteret and S. Lomonaco. It is a precursor to a paper \cite{CKL} by the three of us on these themes. Some of the
information in the present paper can be found in \cite{BG} related to the universality of braiding gates. The paper \cite{Coecke} presents
views about teleportation that are quite similar to ours, and we acknowledge fruitful conversations with Robert Coecke after an early draft of the
present paper was distributed. The references 
\cite{GC,QCJP,TEQE,Spie,KP,L,Lomonaco1,RB,nielsen} are included for background to the present paper.
\bigbreak

The methods of the present paper are motivated by topological considerations, as described in Section 2. Section 2 begins with topology
and shows how it leads to the natural association of a two-by-two matrix with a two-qubit state, and how this association is
related to calculation of quantum amplitudes and to the transmission of quantum information. Section 3 applies these ideas to the 
quantum computation of the absolute value of the trace of a unitary matrix. Section 4 gives applications to teleportation of states and teleportation
of unitary transformations.
\bigbreak

\noindent {\bf Acknowledgement.} Most of this effort was sponsored by the Defense
Advanced Research Projects Agency (DARPA) and Air Force Research Laboratory, Air
Force Materiel Command, USAF, under agreement F30602-01-2-05022. Some of this
effort was also sponsored by the National Institute for Standards and Technology
(NIST). The U.S. Government is authorized to reproduce and distribute reprints
for Government purposes notwithstanding any copyright annotations thereon. The
views and conclusions contained herein are those of the authors and should not be
interpreted as necessarily representing the official policies or endorsements,
either expressed or implied, of the Defense Advanced Research Projects Agency,
the Air Force Research Laboratory, or the U.S. Government. (Copyright 2005.) 
It gives the first author great pleasure to acknowledge support from NSF Grant DMS-0245588,
and to give thanks to the
University of Waterloo and the Perimeter Institute in Waterloo, Canada for their hospitality during
the preparation of this research. We thank Ferando Souza, Sam Lomonaco, Hilary Carteret and Niel de Beaudrap for helpful conversations.
We thank Sergey Kilin for the invitation and opportunity to speak on the ideas in the present paper at the 
Xth International Conference on Quantum Optics 2004 in Minsk, Belarus.
\bigbreak

\section{Topological Amplitudes, States and Measurements}
 In this section we discuss the topology of curves in the plane from the point of view of topological amplitudes.
We think of a minimum as a diagram for the creation of two particles from the vacuum, and a maximum as a diagram for the
annihilation of two particles. We then assign matrices for these creations and annihilations, and calculate the corresponding
mathematical amplitudes. All this can be regarded as a description of how to attach matrices to parts of plane curves in order to 
capture topological properties, but we wish to emphasize the possible quantum physical intepretations. 
\bigbreak

More than one interpretation is possible. We can also regard the minima as standing for state preparations and the maxima as standing for 
measurements. A measurement is different from an annihilation. When we measure a state, we get one of a range of possible
outcomes. When we calculate the amplitudes for annihilation in these matrix situations we are actually calculating the amplitude for a 
successful measurement with respect to a chosen state. Thus for applications to quantum information theory we have to take great care \
in the bookkeeping for the measurements. This will be discussed in detail in the next section.
\bigbreak
 
Consider first a circle in a spacetime plane with time represented vertically and
space horizontally. The circle represents a vacuum to vacuum process that
includes the creation of
two ``particles", and their subsequent annihilation. See Figures 1 and 2.
\bigbreak

{\tt    \setlength{\unitlength}{0.92pt}
\begin{picture}(154,111)
\thicklines   \put(77,60){\circle{40}}
\thinlines    \put(1,78){\makebox(33,32){$t$}}
              \put(127,5){\makebox(26,26){$x$}}
              \put(25,20){\vector(1,0){92}}
              \put(38,1){\vector(0,1){99}}
\end{picture}}

\begin{center}
{\bf Figure 1 - Circle in Spacetime} 
\end{center}
\vspace{3mm}

{\tt    \setlength{\unitlength}{0.92pt}
\begin{picture}(104,132)
\thicklines   \put(3,1){\vector(0,1){130}}
\thinlines    \put(63,44){\oval(80,78)[b]}
              \put(63,84){\oval(80,78)[t]}
\end{picture}}
 
\begin{center}
{\bf Figure 2 - Creation and Annihilation} 
\end{center}
\vspace{3mm}

In accord with our previous description, we could divide the circle into these
two parts (creation(C)  and annihilation (A)) and consider the amplitude   $\langle A|C \rangle .$
Since the diagram for the creation of the two particles ends in two separate
points, it is natural to take a vector space of the form  $V \otimes V$  as the target for
the bra and as the domain of the ket.
\vspace{3mm}

We imagine at least one particle property being catalogued by each dimension of
$V.$  For example, a basis of  $V$  could enumerate the spins of the created
particles.  If $\{e_{a} \}$ is a basis for $V$ then $\{e_{a} \otimes e_{b} \}$ forms a basis for
$V \otimes V.$ The elements of this new basis constitute all possible combinations of the
particle properties. Since such combinations are multiplicative, the tensor product is the
appropriate construction.
\vspace{3mm}

In this language the creation ket is a map  $cup$,

$$cup: C \longrightarrow V \otimes V,$$

\noindent and the annihilation bra is a mapping  $cap$,

$$cap: V \otimes V \longrightarrow C.$$

It is possible to draw a
much more complicated simple closed curve in the plane that is nevertheless
decomposed with respect to the vertical direction into many cups and caps.  In
fact, any simple (no self-intersections) differentiable curve can be rigidly rotated until
it is in general position with respect to the vertical.  It will then be seen to
be decomposed into these minima and maxima.   Our prescriptions for amplitudes
suggest that we regard any such curve as an amplitude via its description as a
mapping from   $C$  to $C$ where $C$ denotes the complex numbers.
\vspace{3mm}

Each simple closed curve gives rise to an amplitude,  but any simple closed curve
in the plane is isotopic to a circle, by the Jordan Curve Theorem.  If these are
topological amplitudes,  then they should all be equal to the original amplitude
for the circle.   Thus the question:  What condition on creation and annihilation
will insure topological amplitudes?  The answer derives from the fact that all
isotopies of the simple closed curves are generated by the cancellation of
adjacent  maxima and minima as illustrated below.
\vspace{3mm}

{\tt    \setlength{\unitlength}{0.92pt}
\begin{picture}(206,181)
\thinlines    \put(203,173){\line(0,-1){172}}
              \put(169,94){\vector(-1,0){45}}
              \put(138,94){\vector(1,0){48}}
              \put(133,107){\line(-1,-1){1}}
              \put(109,103){\line(0,1){77}}
              \put(3,78){\line(0,-1){76}}
              \put(83,101){\oval(54,86)[b]}
              \put(30,76){\oval(54,86)[t]}
\end{picture}}
  
\begin{center}
{\bf Figure 3 - Cancellation of Maxima and Minima} 
\end{center}
\vspace{3mm}

In composing mappings it is necessary to use the identifications 
$(V \otimes V) \otimes V = V \otimes (V \otimes V)$ 
and $V \otimes C = C \otimes V = V.$ Thus in the illustration above, the composition on the left
is given by

$$V = V \otimes C  - 1 \otimes cup \rightarrow  V \otimes (V \otimes V)$$

$$ = (V \otimes V) \otimes V  - cap \otimes 1 \rightarrow  C \otimes V = V.$$

\noindent This composition must equal the identity map on $V$ (denoted $1$ here) for the
amplitudes to have a proper image of the topological cancellation.
This condition is said very simply by taking a matrix representation for the
corresponding operators.
\vspace{3mm}

Specifically,  let   $\{e_{1}, e_{2}, ..., e_{n} \}$  be a basis for $V.$ 
Let $e_{ab} = e_{a} \otimes  e_{b}$  
denote the elements of the tensor basis for $V \otimes V.$  Then there are matrices  $M_{ab}$  
and  $M^{ab}$  such that

$$cup(1)  =  \Sigma M^{ab} e_{ab}$$

\noindent   with the summation taken over all values of $a$ and $b$ 
from $1$ to $n.$  Similarly,  $cap$  is described by 

$$cap(e_{ab}) =  M_{ab.}$$  

\noindent Thus the
amplitude for the circle is 

$$cap[cup(1)]  =  cap \Sigma M^{ab}e_{ab} = \Sigma M^{ab}M_{ab.}$$

\noindent In
general, the value of the amplitude on a simple closed curve is obtained by translating it into an
``abstract tensor expression"  in the $M_{ab}$ and $M^{ab}$, and then summing over these
products for all cases of repeated indices.
\vspace{3mm}

Returning to the topological conditions we see that they are just that the
matrices  $(M_{ab})$  and  $(M^{ab})$  are inverses in the sense that   
$\Sigma M_{ai}M^{ib}  = \delta_{a}^{b}$  and
$\Sigma M^{ai}M_{ib}  = \delta^{a}_{b}$ where $\delta_{a}^{b}$ denotes the (identity matrix)
Kronecker delta that is equal to one when its two indices are equal to one another and zero
otherwise.
\vspace{3mm}

 {\tt    \setlength{\unitlength}{0.92pt}
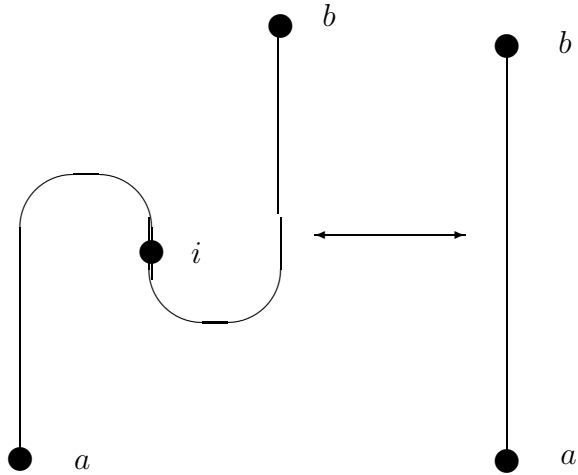
\begin{picture}(247,219)
\thinlines    \put(216,172){\makebox(28,35){$b$}}
              \put(217,3){\makebox(29,31){$a$}}
              \put(119,183){\makebox(28,35){$b$}}
              \put(67,88){\makebox(23,30){$i$}}
              \put(17,1){\makebox(29,31){$a$}}
              \put(206,188){\circle*{10}}
              \put(6,18){\circle*{10}}
              \put(60,103){\circle*{10}}
              \put(113,196){\circle*{10}}
              \put(-24,206){\circle*{0}}
              \put(206,17){\circle*{10}}
              \put(206,189){\line(0,-1){172}}
              \put(172,110){\vector(-1,0){45}}
              \put(141,110){\vector(1,0){48}}
              \put(136,123){\line(-1,-1){1}}
              \put(112,119){\line(0,1){77}}
              \put(6,94){\line(0,-1){76}}
              \put(86,117){\oval(54,86)[b]}
              \put(33,92){\oval(54,86)[t]}
\end{picture}}

\begin{center}
{\bf Figure 4 - Algebraic Cancellation of Maxima and Minima} 
\end{center}
\vspace{3mm}

In Figure 4, we show the diagrammatic representative of  the
equation $\Sigma M_{ai}M^{ib}  = \delta_{a}^{b}.$
\vspace{3mm}

In the simplest case  $cup$ and $cap$  are represented by  $2 \times 2$ matrices.  The
topological condition implies that these matrices are inverses of each other.
Thus the problem of the existence of topological amplitudes is very easily solved
for simple closed curves in the plane.
\vspace{3mm}

Now view Figure 5.

$$ \picill5inby4in(Matrix)  $$

\begin{center}
{\bf Figure 5 - Matrix Composition}
\end{center}

In this Figure we have summarized the essential diagrammatic mathematics of this section.
To a minimum is assigned a matrix $M^{ab}$, and to a maximum is assigned a matrix $M_{ab}.$
A concatenation of a minimum and a maximum gives a linear transformation $N:V \longrightarrow V$
with matrix $$N^{b}_{a} = \Sigma _{i} M_{ai}M^{ib}.$$  The matrix $N$ would be the identity matrix, if we wanted to parallel topological
deformation. In general, $N$ is just the composition of the matrices for the minimum and the maximum.
\bigbreak

We now wish to shift interpretations to states and measurements. To this end, let us associate the state
$$\langle  Cap| = \Sigma_{a,b} M_{a,b}\langle a| \langle b|$$ to the maximum and the state
$$|Cup \rangle  = \Sigma_{a,b} M^{a,b} |a \rangle |b \rangle $$ to the minimum.
Assume for this discussion that the indices range over the values $0$ and $1$ so that we are using $V$ as the space for a single qubit.
We interpret $|Cup \rangle$ as a preparation of a two qubit state. We interpret $\langle  Cap|$ as a two-qubit measurement state.
Let $$| \psi \rangle = \Sigma_{k} \psi_{k} |k \rangle$$ be Alice's initial state starting at the bottom the diagram of
the concatenated maximum and minimum. See Figure 6. Alice tensors $| \psi \rangle$ with $|Cup \rangle$ and then uses $\langle  Cap|$ in a 
successful measurement in the first two tensor factors of $| \psi \rangle |Cup \rangle.$ The state resulting from this successful
measurement can be regarded as Bob's state at the top of the diagram (Alice tells Bob of her success by a classical channel.).
The resulting state is then calculated thus: Note that $\langle i|j \rangle = \delta_{i,j}$ is the Kronecker delta, equal to one when
$i=j$ and equal to zero otherwise.

$$\langle  Cap|(| \psi \rangle |Cup \rangle) $$
$$= \Sigma_{a,j} M_{a,j}\langle a| \langle j| (\Sigma_{k,i,b}\psi_{k}M^{i,b}|k \rangle |i\rangle |b \rangle) $$
$$= \Sigma_{a,j, k, i ,b} M_{a,j}  \psi_{k}M^{i,b}\langle a|k \rangle \langle j|i\rangle |b \rangle $$
$$= \Sigma_{k,i,b} \psi_{k} M_{k,i} M^{i,b} |b \rangle = \Sigma_{k, b} \psi_{k} N_{k}^{b} |b \rangle = N(\Sigma_{k} \psi_{k}|k \rangle) = N |\psi
\rangle. $$

The upshot is that the state transmitted to Bob by this process is $N|\psi \rangle$ where $N$ is the composition of the matrices 
corresponding to the preparation state $| Cup \rangle$ and the measurement state $\langle Cap |.$ This tells us that if we had wanted 
Bob to receive directly a copy of $| \psi \rangle,$ then we would need the matrix for the preparation state $| Cup \rangle$ to be 
invertible. The reader should note that {\it the condition for the invertibility of the matrix associated with $| Cup \rangle$
is exactly equivalent to the condition that this two-qubit state be entangled (not a decomposable tensor product).} 
\bigbreak

$$ \picill5inby5in(Infomatrix)  $$

\begin{center}
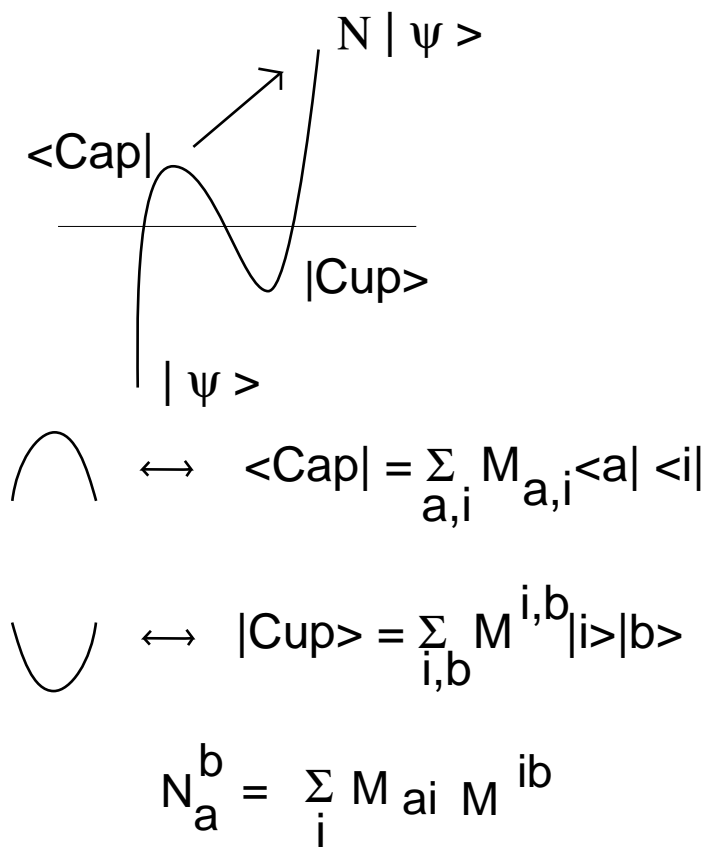
{\bf Figure 6 - Matrix Composition for Preparation and Measurement}
\end{center}

In the next two sections, we use these ideas to analyse the trace of a unitary transformation and the structure of teleportation.
\bigbreak

\section{Trace}

The formalism of configuring a computation in terms of preparation and measurement 
can be used in very general quantum computational contexts. For example, let $U$ be a unitary transformation on
$H = V^{\otimes n}$ where $V$ is the complex two-dimesional space for a single qubit. Represent $U$ as a box with
$n$ input lines at the bottom and $n$ output lines at the top, each line corresponding to a single qubit in an
element of the tensor product $H$ with basis $\{|\alpha\rangle | \alpha \,\, \mbox{is a binary string of length} \, n \}.$ 
Let $|\delta\rangle = \Sigma_{\alpha} |\alpha,\alpha\rangle \in H \otimes H$ where $\alpha$ runs over all binary strings of length
$n.$ Note that $\langle \delta|$ is the following covector mapping $H \otimes H$ to the complex numbers $C:$
$$\langle\delta|\alpha, \beta\rangle = 1 \,\, \mbox{if} \,\, \alpha = \beta \,\, \mbox{and} \,\, \langle\delta|\alpha, \beta\rangle = 0
\,\,
\mbox{otherwise.}$$ Now let $W = U \otimes I_{H},$ where $I_{H}$ denotes the identity transformation of $H$ to $H.$
Then $$\langle\delta|W|\delta\rangle = \langle\delta|U\otimes I_{H}|\delta\rangle = $$ 
$$\langle\delta| \Sigma_{\gamma} U^{\gamma}_{\alpha} |\gamma, \alpha\rangle = \Sigma_{\alpha} U^{\alpha}_{\alpha} = tr(U).$$
For example, $\langle\delta|\delta\rangle = 2^{n} = tr(I_{H}).$ See Figure 7 for an illustration of this process.
\bigbreak

{\tt    \setlength{\unitlength}{0.92pt}
\begin{picture}(334,247)
\thinlines    \put(39,105){\makebox(49,39){$U$}}
              \put(232,105){\makebox(101,40){$U \otimes I_{H}$}}
              \put(221,204){\makebox(81,40){$\langle\delta|$}}
              \put(220,1){\makebox(82,43){$|\delta\rangle$}}
              \put(1,69){\dashbox{6}(210,114){}}
              \put(190,4){\line(-1,0){149}}
              \put(190,243){\line(0,-1){239}}
              \put(39,244){\line(1,0){151}}
              \put(158,24){\line(-1,0){97}}
              \put(159,222){\line(0,-1){198}}
              \put(61,223){\line(1,0){98}}
              \put(124,44){\line(-1,0){44}}
              \put(123,204){\line(0,-1){160}}
              \put(82,204){\line(1,0){41}}
              \put(41,84){\line(0,-1){80}}
              \put(61,84){\line(0,-1){59}}
              \put(80,84){\line(0,-1){41}}
              \put(82,165){\line(0,1){39}}
              \put(61,164){\line(0,1){60}}
              \put(39,163){\line(0,1){81}}
              \put(21,83){\framebox(81,81){}}
\end{picture}}

\begin{center}
{\bf Figure 7 - A quantum process to obtain $|tr(U)|.$} 
\end{center}
\vspace{3mm}

Thus we see that we can, for any unitary matrix
$U,$ produce a quantum computational process with preparation $|\delta\rangle$ and measurement $\langle\delta|$ such that {\em the 
amplitude of this process is the trace of the matrix $U$ divided by $(\sqrt{2})^{n}$.} This means that the corresponding quantum computer
computes the probability associated with this amplitude. This probability is the absolute square of the
amplitude and so the quantum computer will have $|tr(U)|^{2}/2^{n}$ as the probability of success  and hence one can find $|tr(U)|$ by
successive trials. We have  proved the 
\bigbreak

\noindent {\bf Lemma.} With the above notation, the absolute value of the trace of a unitary matrix $U$, $|tr(U)|$,
can be  obtained to any desired degree of accuracy from the quantum computer corresponding to $U \otimes I_{H}$ with preparation
$|\delta\rangle$ and measurement $\langle\delta|,$ where $|\delta\rangle = \Sigma_{\alpha} |\alpha,\alpha\rangle \in H \otimes H.$
\bigbreak

\noindent The proof of the Lemma is in the discussion above its statement.

\section{Teleportation}
The formalism we used at the end of the previous section to describe the (absolute value) of the trace of a unitary matrix contains a
hidden teleportation. It is the purpose of this section to bring forth that hidden connection and to show how this structure illuminates 
the concept of teleportation and its generalizations. 
\bigbreak

First consider the state 
$$|\delta\rangle = \Sigma_{\alpha} |\alpha,\alpha\rangle \in H \otimes H.$$
from the last section, where $H = V^{\otimes n}$  and $V$ is a single-qubit space. 
One can regard $|\delta\rangle$ as a generalization of the $EPR$ state $|00 \rangle + |11 \rangle.$
\bigbreak

Let $|\psi \rangle \in H$ be an arbitrary pure state in $H.$ Let $\langle {\cal M}|$ be an abitrary element of the dual of $H \otimes H$
and consider the possibility of a successful measurement via $\langle {\cal M}|$ in the first two tensor factors of 
$$|\psi \rangle |\delta\rangle \in H \otimes H \otimes H.$$ The resulting state from this measurement will be
$$\langle {\cal M}|[|\psi \rangle |\delta\rangle].$$
If 
$$\langle {\cal M}| = \Sigma_{\alpha, \beta} M_{\alpha, \beta}\langle \alpha | \langle \beta |, $$ then
$$\langle {\cal M}|[|\psi \rangle |\delta\rangle] = 
\Sigma_{\alpha, \beta} M_{\alpha, \beta}\langle \alpha | \langle \beta | \Sigma_{\gamma, \lambda}\psi_{\gamma} |\gamma\rangle
|\lambda \rangle |\lambda \rangle$$
$$= \Sigma_{\alpha, \beta} M_{\alpha, \beta}  \Sigma_{\gamma, \lambda}\psi_{\gamma} \langle \alpha |\gamma\rangle
\langle \beta |\lambda \rangle |\lambda \rangle$$
$$= \Sigma_{\alpha,\beta} M_{\alpha, \beta} \psi_{\alpha} |\beta \rangle$$
$$= \Sigma_{\beta} [\Sigma_{\alpha} M_{\alpha, \beta} \psi_{\alpha}] |\beta \rangle$$
$$= \Sigma_{\beta} (M\psi)_{\beta} |\beta \rangle$$
$$= M |\psi \rangle.$$
Thus we have proved the 
\bigbreak

\noindent {\bf Teleportation Lemma.} Successful measurement via $\langle {\cal M}|$ in the first two tensor factors of 
$$|\psi \rangle |\delta\rangle \in H \otimes H \otimes H$$ results in the state $M | \psi \rangle$ where the matrix $M$ represents the 
measurment state $\langle {\cal M}|$ in the sense that $$\langle {\cal M}| = \Sigma_{\alpha, \beta} M_{\alpha, \beta}\langle \alpha |
\langle \beta |.$$
\bigbreak

\noindent {\bf Remark.} The reader should note that while we have proved this lemma by making the calculation quite explicit in terms 
of matrix indices, the lemma follows at once by using the diagrammatic conventions of Figure 5.
\bigbreak

This Lemma contains the key to teleportation. Let $| \psi \rangle$ be a state held by Alice, where Alice and Bob share the 
generalized $EPR$ state $| \delta \rangle.$ Alice measures the combined state $|\psi \rangle |\delta\rangle$ and reports to Bob
that she has succeeded in measuring via $\langle {\cal M}|$ (from some list of shared transformations that they have in common) by a
classical transmission of information. By the Lemma, Bob knows that he now has access to the state $M |\psi \rangle.$ In this 
generalized version of teleportation, we imagine that Alice and Bob have a shared collection of matrices $M$, each coded by 
a bit-string that can be transmitted in a classical channel. By convention, Alice and Bob might take the zero bit-string to denote
lack of success in measuring in one of the desired matrices. Then Alice can send Bob by the classical channel the information of success
in one of the matrices, or failure. For success, Bob knows the identity of the resulting state without measuring it. See Figure 8
for a schematic of this process.
\bigbreak

$$ \picill4inby2.5in(Teleport)  $$

\begin{center}
{\bf Figure 8 - Matrix Teleportation}
\end{center}

In the case of success, and if the matrix $M$ is unitary, Bob can apply $M^{-1}$ to the transmitted state and know that he now has
the original state $| \psi \rangle$ itself. The usual teleportation scenario, is actually based on a list of unitary transformations
sufficent to form a basis for the measurement states. Lets recall how this comes about.
\bigbreak

First take the case where $M$ is a unitary $2 \times 2$ matrix and let $\sigma_1, \sigma_2, \sigma_3$ be the three Pauli matrices
$$\sigma_{1} =  \left[
\begin{array}{cc}
     0 & 1  \\
     1 & 0
\end{array}
\right] ,\sigma_{2} =  \left[
\begin{array}{cc}
     0 & -i  \\
     i & 0
\end{array}
\right],\sigma_{3} =  \left[
\begin{array}{cc}
     1 & 0  \\
     0 & -1
\end{array}
\right] $$
We replace $\sigma_{2}$ by $i\sigma_{2}$ (for ease of calculation) and obtain the three matrices $X$, $Y$, $Z:$
$$X =  \left[
\begin{array}{cc}
     0 & 1  \\
     1 & 0
\end{array}
\right], Y=  \left[
\begin{array}{cc}
     0 & 1  \\
    -1 & 0
\end{array}
\right],Z =  \left[
\begin{array}{cc}
     1 & 0  \\
     0 & -1
\end{array}
\right] $$
\bigbreak

\noindent {\bf Basis Lemma.} Let $M$ be a $2 \times 2$ matrix with complex entries. Let the {\em measuring state for $M$} be the
state
$$\langle {\cal M}| = M_{00}|00 \rangle + M_{01}|01 \rangle + M_{10}|10 \rangle + M_{11}|11 \rangle.$$
Let $\langle { \cal XM}|$ denote the measuring state for the matrix $XM$ (similarly for $YM$ and $ZM$).
Then the vectors $$\{ \langle {\cal M}|, \langle { \cal XM}|, \langle {\cal YM}|,\langle { \cal ZM}| \}$$
are orthogonal in the complex vector space $V \otimes V$ if and only if $M$ is a multiple of a unitary matrix $U$ of the form
$$U =  \left[
\begin{array}{cc}
     z & w  \\
     -\bar{w} & \bar{z}
\end{array}
\right]$$ with complex numbers $z$ and $w$ as generating entries.
\bigbreak

\noindent {\bf Proof.} We leave the proof of this Lemma to the reader. It is a straightforward calculation.
\bigbreak

This Lemma contains standard teleportation procedure when one takes $M = I$ to be the identity matrix. Then the four measurement states
$$\{ \langle {\cal I}|, \langle { \cal X}|, \langle {\cal Y}|,\langle { \cal Z}| \}$$ form an orthogonal basis and by the Telportation
Lemma, they successfully transmit $\{ |\psi \rangle, X|\psi \rangle, Y|\psi \rangle ,Z|\psi \rangle \}$ respectively. Bob can rotate each
of  these received states back to $|\psi \rangle$ by a unitary transformation (Remember that states are determined up to phase.). In this
form, the Lemma shows that we can, in fact, teleport any $2 \times 2$ unitary matrix transformation $U.$ We take $M = U,$ and take the
othogonal  basis provided by the Lemma. Then a $2$-qubit classical transmission from Alice to Bob will enable Bob to identify the 
measured state and he can rotate it back to $U|\psi \rangle.$

Note that for $H = V^{\otimes n}$ we can consider the matrices $$T_{\alpha, \beta} = T_{\alpha(1),\beta(1)}\otimes \cdots \otimes
T_{\alpha(n), \beta(n)}$$ where $\alpha = (\alpha(1), \cdots , \alpha(n))$ and $\beta = (\beta(1), \cdots , \beta(n))$ are bit-strings
of length $n$ and 
$T_{0,0} = I, T_{0,1} = X, T_{1,0} = Y, T_{1,1} = Z$ are the modified Pauli matrices discussed above. Then just as in the above Lemma, if
$U$ is a  unitary matrix defined on $H,$ then the set of measurement states $\langle {\cal T_{\alpha,\beta}U}|$ for the matrices
$T_{\alpha,\beta}U$ are an orthogonal basis for $H \otimes H.$ Hence we can teleport the action of the arbitrary unitary matrix $U$ from
Alice to Bob, at the expense of a transmission of $2^{n}$ classical bits of information. This means that, we can construct an
arbitrary unitary transformation (hence an idealized quantum computer) almost entirely by using quantum measurments. This result should
be  compared with the results of \cite{GC}, and \cite{RB}, which we shall do in a paper subsequent to the present work.  If Alice and Bob
conicide as observers, then there is no need to transmit the classical bits. The result of a given measurement is an instruction to
perform one of a preselected collection of unitary transformations on the resulting state. 
\bigbreak

\end{document}